\documentstyle[12pt]{article}
\newlength{\myleftmargin}
\newlength{\paperwidth}
\newcommand{\an}{ans{\" a}tze}

\newcommand{\rruuc}{R_u^{uc}}
\newcommand{\rruct}{R_u^{ct}}
\newcommand{\ruuc}{r_u^{uc}}
\newcommand{\ruct}{r_u^{ct}}
\newcommand{\rrdds}{R_d^{ds}}
\newcommand{\rrdsb}{R_d^{sb}}
\newcommand{\rdds}{r_d^{ds}}
\newcommand{\rdsb}{r_d^{sb}}
\newcommand{\ccc}{\circ}
\begin{document}
\thispagestyle{empty}

\begin{flushright}
DPNU-96-07\\
{F}ebruary 1996
\end{flushright}

\vspace{1em}

\begin{center}
{\LARGE{\bf The Unitarity Triangle and}}
\end{center}
\begin{center}
{\LARGE{\bf Quark Mass Matrices on the NNI Basis}}
\end {center}

\vspace{2em}

\begin{center}
{\sc Toshiaki~~Ito}\\
\sl{Department of Physics, Nagoya University,} {\it Nagoya 464-01, Japan}
\end{center}
\begin{center}
{\sc and}
\end{center}
\begin{center}
{\sc Morimitsu~~Tanimoto}\\
\sl{Science Education Laboratory, Ehime University,} {\it Matsuyama 790, Japan}
\end{center}

\vspace{4em}

{\large{\bf Abstract}}

We examine the unitarity triangle of the KM matrix by using 
  the general quark mass matrices on the NNI basis.
  The Fritzsch anz\"atz is modified  by introducing four additional
  parameters.
The KM matrix elements are expressed
in terms of quark mass ratios, two phases and four additional parameters.
 It is found that the vertex of the unitarity triangle  is predicted 
 to be almost  in the second quadrant on the $\rho -\eta$ plane
 as far  as $V_{us}\simeq -{\displaystyle \sqrt{\frac{m_d}{m_s}}e^{ip}+
  \sqrt{\frac{m_u}{m_c}}e^{iq}}$ is held.

\clearpage

{\bf {\S}1. Introduction}

\vspace{1em}

One of the most important unsolved problems of the flavor physics is the
 understanding of the  flavor mixing and  the fermion masses, which
  are free parameters in the standard model.
 The observed values of those mixing and masses provide us
 clues of the origin of the fermion mass matrices.
  The one of the most strict method for the consideration of
  quark mass matrices
 is an examination of the so called unitarity triangle of the
 Kobayashi-Maskawa(KM) matrix[1].
At present, 
   the unitarity triangle in the $\rho - \eta$ plane[2]
 is determined by the experimental data of
$B\to X\ell{\bar\nu}_{\ell}$, $\epsilon$ parameter in the neutral $K$
 meson system and $B_d-{\bar B}_d$ mixing.
However, the experimental allowed region  is too wide to determine the
point of the vertex
 in the $\rho -\eta$ plane.
  The unitarity triangle is expected to be determined precisely
in B-factory at KEK and SLAC in the near future. 
On the other hand, one needs experimental information of six quark masses
 to give reliable estimate  of the  KM matrix elements in the quark mass
matrix models.
The recent discovery of the top quark[3,4] provides us the chance of
 the precise study  of quark mass matrices. 
Thus, we are now in the epoch to examine quark mass matrices
 by focusing on the unitarity triangle.

In this paper, we study 
the unitarity triangle of the KM matrix by using
 the  quark mass matrices on
 the nearest-neighbor interactions(NNI) basis[5].
Any up- and down- $3\times 3$ quark mass matrices
 can be always transformed to this basis by a weak-basis transformation.
We  present general discussions of the unitarity triangle 
 in context with quark mass matrices on the NNI basis.
 It is likely that  the vertex of the unitarity triangle
  almost comes on the second quadrant of the $\rho -\eta$ plane.

 In {\S}2, we give the expression of the KM matrix from general
quark mass matrices on the NNI basis.
The KM matrix elements are expressed in terms of quark mass ratios,
two phases and four additional parameters
that stand for  the discrepancy from the Fritzsch {\an}[6].
We show the vertex of  the unitarity triangle on
 the $\rho -\eta$ plane  in  relation to parameters  of the quark mass
matrices in {\S}3.
Summary is given in {\S}4.

\vspace{2em}

{\bf {\S}2.   Quark mass matrix on the NNI basis and KM matrix}

\vspace{1em}

As presented by Branco, Lavoura and Mota, both up and down
 quark mass matrices could be always transformed to the non-Hermitian matrices 
 on the NNI basis 
 by a weak-basis transformation for the three and four generation cases[5].
In this basis,  the KM matrix elements are expressed generally
 in terms of mass matrix parameters due to eight texture zeros.
In particular, phases of the mass matrices can be isolated easily
 as shown later.
The famous Fritzsch {\an} is the special one  of the NNI basis.
Therefore, the discrepancy from the Fritzsch {\an} can be estimated simply
 on this basis.
We begin with discussing the general  quark mass matrices on the NNI basis
as follows:
\begin{equation}
M_u= \left( 
\begin{array}{ccc}
0 & A & 0 \\
B & 0 & C \\
0 & D & E
\end{array}
\right)
~~, \qquad
M_d= \left( 
\begin{array}{ccc}
0 & F & 0 \\
G & 0 & H \\
0 & I & J
\end{array}
\right),
\end{equation}
where $A\sim J$ are $c$-numbers.
The matrices $U_u$, $U_d$ are defined as unitarity matrices which diagonalize
the Hermitian matrices $H_u=M_uM_u^{\dagger}$ and  $H_d=M_dM_d^{\dagger}$:
\begin{equation}
U^{\dagger}_uH_uU_u=D_u~~, \qquad
U^{\dagger}_dH_dU_d=D_d~,
\end{equation}
where
\begin{equation}
D_u= \left(
\begin{array}{ccc}
m_u^2 & 0     & 0     \\
0     & m_c^2 & 0     \\
0     & 0     & m_t^2
\end{array}
\right)
~~, \qquad
D_d= \left(
\begin{array}{ccc}
m_d^2 & 0     & 0     \\
0     & m_s^2 & 0     \\
0     & 0     & m_b^2
\end{array}
\right).
\end{equation}
On the NNI basis, we can extract phases from each quark mass matrix
 by use of the diagonal phase matrices.
Then, we can write
\begin{equation}
U_u=\phi_uO_u~~, \qquad U_d=\phi_dO_d \label{eqn:orth}
\end{equation}
where $\phi_u={\rm diag}.(e^{ip_u},e^{iq_u},1)$,
$\phi_d={\rm diag}.(e^{ip_d},e^{iq_d},1)$ and $O_u$, $O_d$ are
orthogonal matrices.
We define the phase matrix ${\sl\Phi}$:
\begin{eqnarray}
{\sl\Phi} &=& \phi_u^*\phi_d \\
&=&
\left(
\begin{array}{ccc}
e^{ip} & 0      & 0 \\
0      & e^{iq} & 0 \\
0      & 0      & 1
\end{array}
\right),
\end{eqnarray}
where $p=p_d-p_u$ and $q=q_d-q_u$.
So the KM matrix is given by:
\begin{equation} 
V_{KM}=O_u^T{\sl\Phi}O_d \ .
\end{equation}
The Fritzsch {\an} of the quark mass matrix is the simplest one
 on the NNI basis:
\begin{equation}
M_u= \left( 
\begin{array}{ccc}
0   & a_u & 0 \\
a_u & 0   & b_u \\
0   & b_u & c_u
\end{array}
\right)
~~,\qquad
M_d= \left( 
\begin{array}{ccc}
0   & a_d & 0 \\
a_d & 0   & b_d \\
0   & b_d & c_d
\end{array}
\right).
\end{equation}
Although this {\an} is successful for the  $V_{us}$ element as follows:
\begin{equation}
 V_{us}\simeq  -\sqrt{\frac{m_d}{m_s}}e^{ip}+
\sqrt{\frac{m_u}{m_c}}e^{iq}, \label{eqn:fvus}
\end{equation}
it fails for  $V_{cb}$ 
\begin{equation}
V_{cb}\simeq \sqrt{\frac{m_s}{m_b}}e^{iq}-
\sqrt{\frac{m_c}{m_t}} ,
\end{equation}
as far as $m_t\geq 100$GeV.
Now this simplest {\an} has been ruled out since the top-quark mass is
found to be
  larger than $160$GeV[3,4].
On the other hand, another {\an} proposed by Branco et al.[7]:
\begin{equation}
M_u= \left( 
\begin{array}{ccc}
0   & a_u & 0 \\
a_u & 0   & b_u \\
0   & c_u & c_u
\end{array}
\right)
~~, \qquad
M_d= \left( 
\begin{array}{ccc}
0   & a_d & 0 \\
a_d & 0   & b_d \\
0   & c_d & c_d
\end{array}
\right),
\end{equation}
 is successful  not only for the  $V_{us}$ element:
\begin{equation}
V_{us}\simeq 
-\frac{1}{{}^4\sqrt{2}}\sqrt{\frac{m_d}{m_s}}e^{ip}+
\frac{1}{{}^4\sqrt{2}}\sqrt{\frac{m_u}{m_c}}e^{iq} ,
\end{equation}
but also for $V_{cb}$:
\begin{equation}
V_{cb}\simeq  \frac{m_s}{m_b}e^{iq}-\frac{m_c}{m_t} .
\end{equation}
Although this {\an} overcomes the fault of the Fritzsch {\an}, it cannot
reproduce the observed ratio of ${|V_{ub}|}/{|V_{cb}|}$.
So if we describe the unitarity triangle by use of the {\an} by Branco et al.,
the vertex of the triangle is plotted outside
of the experimental allowed region
on the $\rho - \eta$ plane, as shown in Fig.1. 
Here, in order to describe the experimental allowed region,
 we use the following data[8,9],
\begin{eqnarray}
B_K &=& 0.076\pm 0.04, \\
\frac{|V_{ub}|}{|V_{cb}|} &=& 0.08\pm 0.02, \\
x_d &=& 0.19\pm 0.01.
\end{eqnarray}

 Therefore we should consider more general matrices in order to examine
 the unitarity triangle.
   We adopt a weak hypothesis: the generation hierarchy of the matrix elements,
   which was guaranteed in the Fritzsch {\an} due to the observed mass
hierarchy,
\begin{equation}
M_{12}\approx M_{21}\ll M_{23}\approx M_{32}\ll M_{33}.
\end{equation}
The quark mass matrices are set in general as
\begin{equation}
M_u= \left( 
\begin{array}{ccc}
0    & a_u  & 0 \\
a_uz & 0    & b_u \\
0    & b_ux & c_u
\end{array}
\right)
~~,\qquad
M_d= \left( 
\begin{array}{ccc}
0    & a_d  & 0 \\
a_dw & 0    & b_d \\
0    & b_dy & c_d
\end{array}
\right), \label{eqn:nni}
\end{equation}
where $x$, $y$, $z$ and $w$ are parameters that stand for discrepancies
from the Fritzsch mass matrix.
Because of the hierarchy of mass matrix elements,
we may restrict the parameter region:
\begin{equation}
0.1\leq x,y,z,w \leq 10.
\end{equation}
  We  comment on the number of parameters.
  There are twelve parameters in the mass matrices.
 Since the number of the  physical parameters are
  ten(six masses and four KM mixing parameters),
 the basis in Eq.(\ref{eqn:nni})
   is general one.  In order to give predictions, one needs some
   assumption on the parameters of the mass matrices.
   
In order to obtain the elements of $O_{u(d)}$ in Eq.(\ref{eqn:orth}),
 we express $a_{u(d)}$,
$b_{u(d)}$ and $c_{u(d)}$ in terms of quark masses from eigenvalue equations
of $H_{u(d)}$.
{F}or example, the eigenvalue equation for $c_u$ is written as
\begin{equation}
c_u^4-2M_1c_u^2+\frac{(1+x^2)^2(1+z^2)}{x^2z}\sqrt{M_3}c_u+
M_1^2-\frac{(1+x^2)^2}{x^2}M_2=0 \label{eqn:eqcu}
\end{equation}
where
\begin{eqnarray}
M_1 &=& m_u^2+m_c^2+m_t^2, \\
M_2 &=& m_u^2m_c^2+m_c^2m_t^2+m_t^2m_u^2, \\
M_3 &=& m_u^2m_c^2m_t^2.
\end{eqnarray}
Neglecting the third term of Eq.(\ref{eqn:eqcu}), which is very small,
we obtain $c_u$
\begin{equation}
c_u \simeq m_t\left( 1-\frac{1+x^2}{2x}\frac{m_c}{m_t}\right) .
\end{equation}
{F}urthermore, we get
\begin{eqnarray}
a_u &\simeq& \sqrt{\frac{m_um_c}{z}}\left( 1+\frac{1+x^2}{4x}
\frac{m_c}{m_t}\right), \\
b_u &\simeq& \sqrt{\frac{m_cm_t}{x}}\left( 1-\frac{1+z^2}{4z}
\frac{m_u}{m_c}\right).
\end{eqnarray}
Then the matrix elements of $O_u$ are written as
\begin{eqnarray}
(O_u)_{11} &\simeq& 1/N_{u_1}, \\
(O_u)_{12} &\simeq& -\frac{1}{\sqrt{z}}\sqrt{\frac{m_u}{m_c}}(1+R_u^{ct}+
r_u^{ct}+r_u^{uc})/N_{u_2}, \\
(O_u)_{13} &\simeq& \sqrt{\frac{x}{z}}\frac{m_c}{m_t}\sqrt{\frac{m_u}{m_t}}
(1+R_u^{ct}-R_u^{uc})/N_{u_3}, \\
(O_u)_{21} &\simeq& \frac{1}{\sqrt{z}}\sqrt{\frac{m_u}{m_c}}(1-R_u^{ct}
+r_u^{uc})/N_{u_1}, \\
(O_u)_{22} &\simeq& 1/N_{u_2}, \\
(O_u)_{23} &\simeq& \frac{1}{\sqrt{x}}\sqrt{\frac{m_c}{m_t}}(1-R_u^{uc}+
r_u^{ct})/N_{u_3}, \\
(O_u)_{31} &\simeq& -\frac{1}{\sqrt{xz}}\sqrt{\frac{m_u}{m_c}}(1+R_u^{ct}
-R_u^{uc}+r_u^{uc})/N_{u_1}, \\
(O_u)_{32} &\simeq& -\frac{1}{\sqrt{x}}\sqrt{\frac{m_c}{m_t}}(1-R_u^{uc}+
r_u^{ct})/N_{u_2}, \\
(O_u)_{33} &\simeq& 1/N_{u_3}
\end{eqnarray}
where $N_{u_i}(i=1,2,3)$ are
normalization factors and
\begin{eqnarray}
R_u^{uc} &=& \frac{1+z^2}{4z}\frac{m_u}{m_c}~~, \qquad
R_u^{ct} = \frac{1+x^2}{4x}\frac{m_c}{m_t}, \\
r_u^{uc} &=& \frac{1-z^2}{2z}\frac{m_u}{m_c}~~, \qquad
r_u^{ct} = \frac{1-x^2}{2x}\frac{m_c}{m_t}.
\end{eqnarray}
The matrix elements for  $O_d$ are given analogously.
Here we neglect the terms of order of $m_u/m_t$, $m_d/m_b$ and higher
order corrections.
Thus the KM matrix elements are expressible in terms of quark mass ratios,
two phases and four additional parameters $x$, $y$, $z$, $w$:
\begin{eqnarray}
V_{ud} &\simeq& \frac{1}{N_{ud}}e^{ip}, \label{eqn:vud} \\
V_{us} &\simeq& \frac{1}{N_{us}}\left [
-\sqrt{\frac{m_d}{wm_s}}\left(1+\rrdsb 
+\rdsb +\rdds\right) e^{ip} \right. \nonumber \\
&+&
\left. \sqrt{\frac{m_u}{zm_c}}
\left(1-\rruct +\ruuc \right) e^{iq}\right ] , \label{eqn:vus} \\
V_{ub} &\simeq& \frac{1}{N_{ub}}\left [
\sqrt{\frac{m_u}{xzm_t}}
\left( 1+\rruct -\rruuc+\ruuc \right) \right ] , \label{eqn:vub} \\
V_{cd} &\simeq& \frac{1}{N_{cd}}\left [
-\sqrt{\frac{m_u}{zm_c}}\left( 1+\rruct 
+\ruct+\ruuc \right) e^{ip} \right. \nonumber \\
&+&
\left. \sqrt{\frac{m_d}{wm_s}}
\left(1-\rrdsb+\rdsb\right) e^{iq}\right ] , \label{eqn:vcd} \\
V_{cs} &\simeq& \frac{1}{N_{cs}}e^{iq}, \label{eqn:vcs} \\
V_{cb} &\simeq& \frac{1}{N_{cb}}\left [
\sqrt{\frac{m_s}{ym_b}}\left( 1-\rrdsb+\rdsb\right) e^{iq}-
\sqrt{\frac{m_c}{xm_t}}
\left( 1-\rruuc+\ruct\right)\right ] , \label{eqn:vcb} \\
V_{td} &\simeq& \frac{1}{N_{td}}\left [
\sqrt{\frac{m_d}{ywm_b}}
\left(1+\rrdsb-\rrdds+\rdds\right)\right ] , \label{eqn:vtd} \\
V_{ts} &\simeq& \frac{1}{N_{ts}}\left [
\sqrt{\frac{m_c}{xm_t}}\left( 1-\rruuc+\ruct\right) e^{iq}-
\sqrt{\frac{m_s}{ym_b}}\left(1-\rrdds
+\rdsb\right)\right ] , \label{eqn:1vts} \\
V_{tb} &\simeq& \frac{1}{N_{tb}}, \label{eqn:vtb}
\end{eqnarray}
where $N$'s are normalization factors such as $N_{us}=N_{u_1}N_{d_2}$ etc..

\vspace{2em}

{\bf {\S}3.  Unitarity Triangle and Quark Mass Matrix Parameters}

\vspace{1em}

In order to estimate the absolute values of
KM matrix elements $|V_{ij}|$, we must know the values
of the masses of six quarks on a same energy scale.
{F}ollowing the recent study by Koide[10],
 we obtain the values of quark masses
at 1GeV
by use of the 2-loop renormalization group equation:
\begin{eqnarray*}
m_u &=& 0.0056\pm 0.011~~,~m_d=0.0099\pm 0.0011~~,~m_s=0.199\pm 0.033~~, \\
m_c &=& 1.316\pm 0.024~~,~m_b=5.934\pm 0.101~~,
~m_t=349.5\pm 27.9~({\rm GeV})~~,
\end{eqnarray*}
where $\Lambda^{(5)}_{\overline{MS}}=0.195$GeV.

Now we can estimate of KM matrix elements.
At first, we set $z=w=1$, which corresponds to the Fritzsch {\an} for 
the first- and second-generation mixing sector
since this {\an} works well for $|V_{us}|$.
 It is remarked  that the following results are available for all models which
express $|V_{us}|$   as $- {\displaystyle \sqrt{\frac{m_d}{m_s}}e^{ip}+
  \sqrt{\frac{m_u}{m_c}}e^{iq}}$ because our KM matrix elements
 in Eq.(\ref{eqn:vus})
are expressed
   generally.\par
   
Since we examine the unitarity triangle which is normalized by
$V^*_{cb}V_{cd}$, we should choose parameters that reproduce the experimental
values of $|V_{cb}|$ and $|V_{cd}|$.
So we use two observed values[9],
\begin{equation}
|V_{cd}|=0.2196\pm 0.0018~~, \qquad |V_{cb}|=0.040\pm 0.005
\end{equation}
as input parameters.
In the  numerical analyses,
 we use the center values of these experimental data.
  The ambiguity  of our predictions due to  experimental
 errors is rather small.
{F}or given  values of $x$ and $q$, we can obtain the value of parameter $y$
by use of Eq.(\ref{eqn:vcb}).
In order to reproduce $|V_{cb}|$, we cannot set $x=y=1$,
that means the unsuccess of  the Fritzsch {\an}. 
 In Fig.2, the allowed lines  for $|V_{cb}|$ are shown on the  $x-y$ plane
 in the case of $q=0$.
 We find that the phase $q$ is restricted to
$-60^{\ccc}\leq q \leq 60^{\ccc}$ due to the condition $0.1\leq y \leq 10$.
{F}urthermore,  we obtain the value of the phase $p$ from Eq.(\ref{eqn:vcd}).
Since  $p$ is  a real number,
 the value of $x$ is also restricted for each
value of $q$ such as the following examples:
\begin{center}
\begin{tabular}{lcl}
$q = \pm 60^{\ccc}$ &~~~:~~~&    $ 2.0 \leq x \leq 10,$ \\
$q = \pm 30^{\ccc}$ &~~~:~~~&    $ 1.0 \leq x \leq 10,$ \\
$q = 0^{\ccc}$ &~~~:~~~& $0.1\leq x \leq 10.$
\end{tabular}
\end{center}
Thus if $x$ and $q$ are fixed, we can calculate the KM matrix elements.

Let us present the unitarity triangle.
If the phase $q$ is fixed,
the vertex  of the triangle moves on the $\rho -\eta$ plain according
  to the change of parameter $x$.
 We  show  the changes of  the vertices by the dotted lines 
 for fixed value of $q$.
{F}ig.3 shows such lines of the cases $q=-60^{\ccc}$, $-30^{\ccc}$,
$-10^{\ccc}$, $0^{\ccc}$, $10^{\ccc}$,
$30^{\ccc}$ and $60^{\ccc}$.
It is found that  the triangle vertex  is sensitive to the value of $x$
 in the $x\leq 1$ region, but insensitive to $x$ in the large $x$ region.
 If $-60^{\ccc}\leq q\leq 17^{\ccc}$,
a part of the dotted line comes into the experimental allowed region on
the $\rho -\eta$ plane.
The triangle vertices are found to be almost on the second quadrant of the
$\rho - \eta$ plane.
 If  the phase $q$ is in between $-3^{\ccc}$ and $16^{\ccc}$
and $x$ is small,
the triangle vertex is in the experimental allowed region on the first
quadrant.
The regions of $x$ that the vertex comes on the first quadrant
are shown for each $q$ value in the Table 1.
 It is emphasized  that  the vertex of the unitarity triangle is almost
on the second quadrant of the $\rho - \eta$ plane as far as  $z=w=1$,
 namely,  $V_{us}\simeq - {\displaystyle \sqrt{\frac{m_d}{m_s}}e^{ip}+
  \sqrt{\frac{m_u}{m_c}}e^{iq}}$ is held.
  \par

Next, we consider the cases of $z=w\neq 1$.
Due to the condition that $p$ is a real number, 
$z=w\leq 1.3$ is restricted.
 In the region of $1\leq z=w\leq 1.3$, each line of Fig.3 moves down to the
 right side .
So, in these cases, the predicted region of the vertex in the first quadrant,
    becomes wider than the case of $z=w=1$.
While in the region of  $z=w\leq 1$, the each line of Fig.3 moves up to the
left side.
The triangle vertices never come in the allowed region on the
first quadrant in the case of $z=w\leq 0.9$.
{F}ig.4 and 5 show the cases of $z=w=1.3$ and $z=w=0.7$, respectively.
Thus,  we conclude that the vertex of the unitarity
triangle almost appears in the second quadrant on $\rho -\eta$ plane
as far as $z=w$ is assumed.\par

What is the condition that the vertex of the unitarity
triangle  appears in the first quadrant on $\rho -\eta$ plane?
As seen in Eqs.(\ref{eqn:vub}) and (\ref{eqn:vtd}),
 $V_{ub}$ and $V_{td}$ are written roughly as follows:
 
 \begin{equation}
 V_{ub} \simeq  \sqrt{\frac{m_u}{xzm_t}}\ , \qquad
 V_{td} \simeq\sqrt{\frac{m_d}{ywm_b}} \ .
\end{equation}
\noindent
Due to $m_u\sim m_d$ and $m_t\gg m_b$,  the magnitude of $V_{td}$ 
 is expected to be larger than the one of  $V_{ub}$
  for the case of $x\sim y$ and $w\sim z$.
  This fact suggests us that conditions $w>1$ and  $z<1$ are necessary  in
order to move
   the vertex of the unitarity triangle 
    into the first quadrant on $\rho -\eta$ plane.
 Actually, we get  the vertex  
    in the first quadrant on $\rho -\eta$ plane for some cases of $w>1$ and
 $z<1$.
{F}or example, we show one case in Fig.6, in which $x=1.0$, $y=2.8$, $z=0.4$,
$w=2.0$, $p=128.2^{\ccc}$ and $q=0^{\ccc}$ are taken.
  Then, $V_{us}$ is not any longer expressed
 as in the simple form in Eq.(\ref{eqn:fvus}).

\vspace{2em}

{\bf {\S}4. Summary}

\vspace{1em}

In this paper, we examine  the unitarity triangle by using the general
 quark mass matrices on the NNI basis.
{F}or quark mass matrices, we introduce four additional parameters
$x$, $y$, $z$ and $w$, which are restricted $0.1\leq x,y,z,w\leq 10$
due to the assumed generation hierarchy.
Then the KM matrix elements are written in terms of quark mass ratios,
two phases and four parameters.
 In the case of  $z=w$, the vertex of the unitarity triangle
almost comes on the second quadrant of the $\rho -\eta$ plane
as shown in Fig.3, 4 and 5.
If the vertex of the unitarity triangle is determined  in
the neighborhood of the vertex of Fig.6 at $B$-factory,
we we cannot  set $z=w$.
In particular, the quark mass matrix models which predict
  $V_{us}\simeq  - {\displaystyle \sqrt{\frac{m_d}{m_s}}e^{ip}+
  \sqrt{\frac{m_u}{m_c}}e^{iq}}$  will be ruled out.
 Thus, the determination of the unitarity triangle gives impact
  on the  study of the quark mass matrix model.

\vskip 1 cm
\noindent
{\bf Acknowledgments}\par
We would like to thank Prof. Y. Koide for helpful comments.
This research is supported by the Grant-in-Aid for Science Research,
Ministry of Education, Science and Culture, Japan(No. 07640413).

\clearpage

{\large {\bf Table Caption}}


Tab.1 :  The  allowed region of $x$, in which  the vertex of the unitarity
triangle
 is in the first quadrant of the $\rho -\eta$ plane.

\vspace{2em}

{\large {\bf Figure Captions}}


{F}ig.1 :  The unitarity triangle in the {\an} by Branco et al..


{F}ig.2 : The relation between $x$ and $y$ at $q=0^{\ccc}$.
The solid line is the case of $|V_{cb}|=0.040$.
The upper (lower) dashed line is the case of $|V_{cb}|=0.035$
 ($|V_{cb}|=0.045$).


{F}ig.3 : The vertices of the unitarity triangle in the case of $z=w=1$.
Each dotted line is the case of $q=+60^{\ccc}$, $+30^{\ccc}$, $+10^{\ccc}$,
$0^{\ccc}$, $-10^{\ccc}$, $-30^{\ccc}$, $-60^{\ccc}$ from the upper side, 
respectively.
The dots of $q=0^{\ccc}$ and $\pm 10^{\ccc}$ lines correspond to
$x=0.1$, 0.2, 0.4, 0.6, 0.8, 1.0, 2.0, 4.0, 6.0, 8.0, 10.0 in order 
from the right hand side.
The dots of $q=\pm 30^{\ccc}$ lines correspond to
$x=1.0$, 2.0, 3.0, 4.0, 5.0, 6.0, 7.0, 8.0, 9.0, 10.0 in order 
from the right hand side.
The dots of $q=\pm 60^{\ccc}$ lines correspond to
$x=4.0$, 5.0, 6.0, 7.0, 8.0, 9.0, 10.0 in order 
from the right hand side.


{F}ig.4 : The vertices of the unitarity triangle in the case of $z=w=1.3$.
Each dotted line is the case of $q=+60^{\ccc}$, $+30^{\ccc}$, $+10^{\ccc}$,
$0^{\ccc}$, $-10^{\ccc}$, $-30^{\ccc}$ from the upper side, 
respectively.


{F}ig.5 : The vertices of the unitarity triangle in the case of $z=w=0.7$.
Each dotted line is the case of $q=+60^{\ccc}$, $+30^{\ccc}$, $+10^{\ccc}$,
$0^{\ccc}$, $-10^{\ccc}$, $-30^{\ccc}$, $-60^{\ccc}$ from the upper side, 
respectively.


{F}ig.6 : One case in which the vertex of the unitarity triangle comes on the
center of the first quadrant of the $\rho -\eta$ plane.

\clearpage

\begin{center}

{\large{\bf Tab.1}}

\vspace{2em}

\begin{tabular}{|l|l|}
\hline
 values of phase $q$ &  regions of $x$ \\ \hline
$q=-3^{\ccc}$ & $x\leq 0.101$ \\
$q=0^{\ccc}$ & $x\leq 0.108$ \\
$q=5^{\ccc}$ & $x\leq 0.12$ \\
$q=7^{\ccc}$ & $x\leq 0.14$ \\
$q=10^{\ccc}$ & $x\leq 0.18$ \\
$q=12^{\ccc}$ & $x\leq 0.21$ \\
$q=14^{\ccc}$ & $x\leq 0.25$ \\
$q=16^{\ccc}$ & $x\leq 0.31$ \\
\hline
\end{tabular}
\end{center}


\clearpage

\begin{thebibliography}{1}
\bibitem{1}
M. Kobayashi and Maskawa, Prog. Theor. Phys. {\bf 49}, 652(1973).
\bibitem{2}
L. Wolfenstein, Phys. Rev. Lett. {\bf 51}, 1945(1983).
\bibitem{3}
CDF Collaboration, F. Abe et al., Phys. Rev. Lett. {\bf 73}, 225(1994).
\bibitem{4}
D0 Collaboration, S. Abachi et al., Phys. Rev. Lett. {\bf 74}, 2632(1995).
\bibitem{5}
G. C. Branco, L. Lavoura and F. Mota, Phys. Rev. {\bf D39}, 3443(1989).
\bibitem{6}
H. Fritzsch, Phys. Lett. {\bf 73B}, 317(1978); Nucl. Phys. {\bf B115},
189(1979).
\bibitem{7}
G. C. Branco  and J.I. Silva-Marcos, Phys. Lett. {\bf 331B}, 390(1994).
\bibitem{8}
S. Hashimoto(JLQCD Collaboration), \\
Talk at the meeting of Particle Physics in future, YITP, 19 January 1996.
\bibitem{9}
Particle Data Group, L. Montanet et al., Phys. Rev. {\bf D50}, 1173(1994).
\bibitem{10}
Y. Koide, preprint in University of Shizuoka, US-94-05(1994)(unpublished).
\end{thebibliography}
\end{document}